# Interplay between Phase Transformation Instabilities and Spatiotemporal Reaction Heterogeneities in Particulate Intercalation Electrodes

*Shubham Agrawal[1] and Peng Bai[1,2,]**

[1]Department of Energy, Environmental & Chemical Engineering, Washington University in St. Louis, St. Louis, Missouri – 63130, United States

[2]Institue of Materials Science and Engineering, Washington University in St. Louis, St. Louis, Missouri – 63130, United States

*Email: pbai@wustl.edu

# Abstract

Lithium-ion batteries rely on particulate porous electrodes to realize high performance, especially the fast-charging capability. To minimize the particle-wise reaction heterogeneities that may lead to local hot spots, deeper understandings of these electrodes at the mesoscale, i.e. hundreds of particles, have become an urgent need. This study reveals that the seemingly random reaction heterogeneities are actually controlled by the interplay between the non-equilibrium material thermodynamics and the external electrochemical driving force. Our *operando* experiments confirm the true working current density around a single particle that is much higher than the globally averaged current density, can change the behavior of phase transformation. The combined theoretical and experimental analyses reveal that unlike other phase-transforming porous electrodes, not all phase separation processes in graphite can be suppressed at high currents, due to the characteristics of the concentration-dependent exchange current density. The insights highlight the necessity to incorporate materials thermodynamics into electrochemical models to ensure self-consistent understandings of practical porous electrodes toward precision design and management.

# 1. Introduction

The great success of lithium-ion batteries (LIBs) has allowed them to penetrate the markets of consumer electronics and electric vehicles.[1–3] The dynamic performance, cycle life, and safety of LIBs are directly dependent on the microscopic heterogeneities arisen from multiple fully-coupled electrochemical dynamic processes, which however are still not completely understood.[4,5] The state-of-charge (SOC) heterogeneities or non-uniform distribution of charge in both the phase-transforming[6–9] and solid solution[10–14] materials occur due to either the far-from-equilibrium thermodynamics of the active material or the heterogeneous microstructure of the composite electrode. In solid-solution cathode materials in particular,[11,12,15] SOC heterogeneities were commonly attributed to structural heterogeneities, such as the non-uniform distributions of conductive additives, electrolyte, etc., but the actual root cause is worth careful examination, as the well-known solid-solution materials may exhibit "fictitious phase transformation" only under fast-rate conditions.[16] The SOC heterogeneities in battery electrodes also influence the accurate estimation of battery SOC in the battery management systems (BMS) which are responsible for their safe and reliable operation, prevention of physical damages, and handling of thermal degradation and cell unbalancing.[17–19] Sophisticated *operando* techniques employing synchrotron-based X-ray,[11,13,15,20–24] lasers[12,14] and visible light,[6,25–28] have been developed to generate valuable insights and obtain quantitative understandings about the effect of reaction heterogeneities on the intercalation mechanism and degradation behavior in the battery materials. Among them, the *operando* characterization using benchtop optical microscopes have demonstrated a well-rounded balance of large field of view with hundreds of particles, sub-micron resolution, fast imaging rate, and accurate interpretation of the electrochemical response.[12,14,25,27,28] This economical platform facilitates comprehensive and thorough examination of widely used electrode materials to gain deeper understandings of the reaction heterogeneities.

Graphite is in general one of the most studied carbon materials and is the dominant anode material for lithium-ion batteries, yet can be used as cathode in other battery systems.[29–32] Unlike other reversible electrode materials, graphite experiences multiple phase transformation steps upon ion intercalations, also known as ordered stages.[33–35] During lithiation into graphite, the optical characterization of phase transformation reveals three major stages visible as blue (Stage 3), red (Stage 2), and gold (Stage 1) colors.[28,36] However, the widely accepted theory that the phase transformations can be suppressed[37] has not been investigated systematically in graphite,[36,38] due to the lack of cohesive combination of *operando* characterization tools and mathematical analyses. The mechanism of phase transformation in graphite affects its rate capability and cycle life and needs a careful inspection to ensure improved performance.

Here, we use our benchtop *operando* platform to observe the reaction pathway of graphite (de)lithiation in real-time under both slow and fast galvanostatic conditions. This observation allows us to determine the true local current density, which is at least 3 orders of magnitude higher than the average current density over the Brunauer-Emmett-Teller (BET) specific surface area and is the true driving force that governs the phase separation dynamics within individual graphite particles.[28] The linear stability analysis, based on the non-equilibrium thermodynamics, verified by the true local current densities offers clear physical explanations to the suppression or persistence of phase separation in graphite particles. With the phase-field simulation of 200 particles, the insights obtained from this study not only can help optimize other advanced *operando* characterizations strategies including those based on synchrotron X-ray, but also can help improve operation protocols of battery charging and discharging to minimize the dynamic heterogeneities toward longer battery cycle life.

## 2. Results

Our simple optical method has the capability to reveal the subtleties of $Li^+$ ion (de)intercalation into practical graphite electrode under different galvanostatic conditions, allowing us to capture the true localized electrochemical kinetics. While the phase transformation dynamics may seem random initially, it is governed by the non-equilibrium thermodynamics of the material, often overlooked in the literature, yet are different from other phase-transforming materials like $LiFePO_4$.

## 2.1. Phase transformations in *operando* galvanostatic cycling

We first evaluated the transition of colors in graphite particles from grey to blue, red and gold, under slow (0.1 C) and fast (1 C) galvanostatic conditions between 0.0004 V and 1.5 V vs Li/Li$^+$ in Li-graphite half cells (**Figure 1 and Video S1-S4**). At the critical SOCs shown in Fig. 1 (a), all particles in the porous electrode appeared the same color, indicating the same stage, but the dynamic phase transformation processes between the critical stages were highly heterogeneous. The entire lithiation process followed a sequential phase transformation from empty to stage 3 (grey to blue, Step I), stage 3 to stage 2 (blue to red, Step II), and stage 2 to stage 1 (red to gold, Step III), which was reversed during delithiation. Step I of The initial grey to blue transition always follows a solid-solution reaction pathway, during which all particles react concurrently. To examine the intricacies of the two phase-transformation steps (II and III) closely, we selected five representative particles, P1 – P5, highlighted in **Figure 1 (a)** and magnified in **Figure 1 (b) – (d)**.

During 0.1 C lithiation (**Video S1**), the new phases in both phase transformation steps always nucleate at the particle edge to create sharp phase boundaries, which sweep across the particle to complete the phase transformation. However, at a large scale, the intra-particle blue-red transformation is associated with a highly random and selective inter-particle dynamics (**Video S5**). Like the particle-by-particle intercalation in LiFePO$_4$ electrode at a low current,[20] the red phase nucleated in particle P3 first, followed by P1. Particles P2, P4, and P5 remained idle (blue) until P3 and P1 became completely red (**Figure 1(b)**). Moments later, the red phase nucleated in particle P2 and took nearly an hour to complete the transformation, whereas the smaller particles P4 and P5 easily transformed into red color, successively, due to smaller volume. A similar sequential blue-red transformation occurred in all the other particles, suggesting that only a limited fraction of the electrode was "active" at any instant. The red-to-gold phase transformation at 0.1 C current, however, started simultaneously in most of the particles (**Video S5**), within which sharp phase boundaries between the red and gold colors were still clearly visible. More obvious than in the blue-to-red transformation, smaller particles such as P4 and P5 incubated successful nucleations much sooner and got fully-filled much earlier than larger particles. The delithiation process at 0.1 C current triggered the random nucleations of the red phase in most gold particles, essentially reversed the lithiation process until SOC reaches 55%, as shown in **Figure 1 (c)** and **Video S2**. The red-to-blue phase transformation reflected by the particles in the **Video S6** and last four columns of **Figure 1 (c)**, however, did not follow the reversed pathway of particle-by-particle lithiation. Instead, it behaved more like the gold-to-red transformation with most particles experiencing the phase transformation concurrently. Despite the subtle differences, it is evident that the phase transformations in Steps II and III occur through clear phase separation at slow galvanostatic condition during both lithiation and delithiation.

During the lithiation at 1 C current, the blue-to-red transformation appeared more homogeneous than in the case of 0.1 C current, showing smeared phase boundaries and cloudy domains and thus, resembling a solid-solution mechanism, as visible in **Videos S3 and S5** and 22% - 39% SOCs in **Figure 1 (d)**. This observation is consistent with the suppression of phase separation predicted[37] and confirmed[20,39] in $LiFPO_4$ electrodes. However, the red-to-gold transformation remained unaffected by the elevated current density, random nucleations of the gold phase at the edges of multiple particles generated sharp phase boundaries that swept across the particles to complete the phase transformation, just like what we observed at 0.1 C current. The high reaction overpotential at 1 C current, shown in **Figure S1**, resulted in only 65% of the total capacity at the cut-off voltage, leaving a few particles partially lithiated, still with clear red-gold phase boundaries. These active phase boundaries within partially lithiated particles, visible in **Video S4**, may recede to disappear via inter-particle exchanges during a long time relaxation.[39] In our experiments, they started to move with newly nucleated phase boundary during the subsequent delithiation at 1C current, until all particles turned red. Unlike the rather homogeneous blue-to-red lithiation at 1 C current, the red-to-blue delithiation process at 1C always induced random nucleations and two-phase coexistence, as can be seen in the last four columns of **Figure 1 (d)** and **Video S6**.

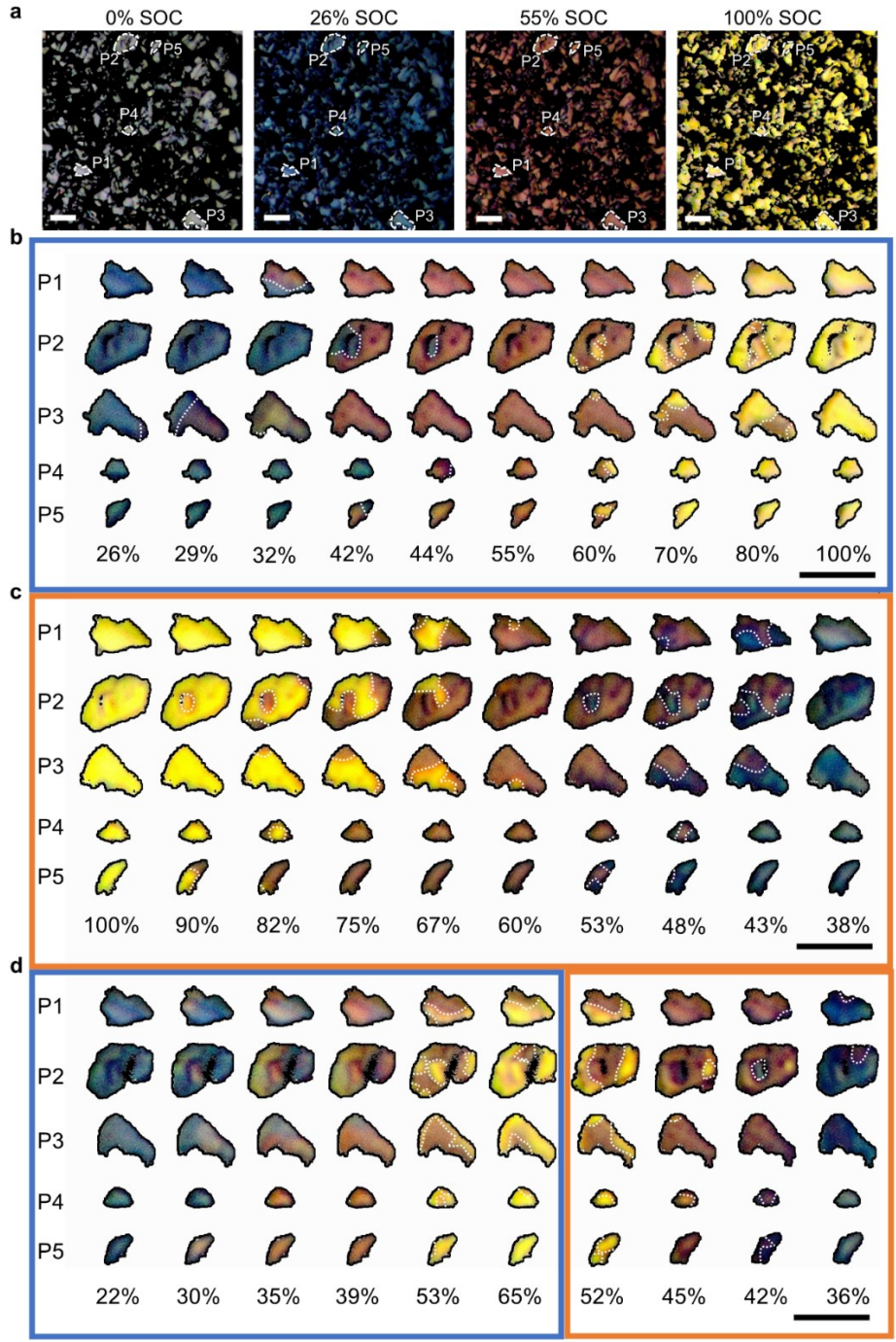

**Figure 1: Phase-transformation during (de)lithiation in graphite under constant current.** (a) Entire view of graphite electrode under the optical microscope in the empty state (grey), and stages 3 (blue), 2 (red), and 1 (gold). These frames are obtained at very low current (C/72) and are used as the calibration frames for the estimation of the SOCs of each stable phase. The white dashed outlines in these snapshots are the five selected particles (P1 – P5) for intra-particle inspection. (b) Lithiation process of the selected

five particles at 0.1 C current. (c) Delithiation of the selected five particles at 0.1 C current. (d) Lithiation and delithiation of the selected five particles at 1 C current. The blue rectangle indicates lithiation while the orange rectangle denotes delithiation. The white dotted lines in the panels (b) – (d) indicate the phase boundaries generated during the phase transformations. The lithiation and delithiation processes at 0.1 C current show the phase transformations along the generated phase boundaries while the lithiation process at 1 C from 22% - 39% SOCs shows smeared phase boundaries and cloudy domains, resembling a solid solution mechanism. *Scale bar: 10 μm.*

The above observation raises two intriguing questions: (i) why does the highly reversible graphite electrode show asymmetrical dynamics of phase transformation during lithiation and delithiation? (ii) Why the high current can suppress the phase separation in the blue-to-red transformation, but not the red-to-gold transformation?

## 2.2. True local electrochemical driving force

Thanks to the sufficient resolution of the distinct colors, we propose to determine the local SOC by converting the color to the standard capacity associated with the color (stage), and calibrated our analysis as explained in **Experimental Section** and **Figure S4**. Following the same principle validated at the larger scale, we calculated the capacity evolution within the selected particles P1 – P5, as shown in **Figure S5**. In the solid-solution regimes, the capacities increased/decreased linearly in each particle, yielding low constant currents. However, in the phase-separation regimes, the total capacity within a particle was the net capacity contributed by each color. The total capacity remained constant when the particle was idling. We obtained the temporal variation of the current carried by each particle by taking the numerical first-order time-derivative of the capacity evolution curves, shown in **Figure S5**. These local "working" currents represent the true electrochemical conditions experienced by each particle, under the global slow and fast galvanostatic conditions.

To determine the local working current *density* for consistent transport analysis in each particle, the active reaction interfaces need to be identified in addition to the current. In general, Li$^+$ ions intercalate through all edge planes of the graphite particle and not the basal planes. The active reaction area is the product of the length of the perimeter and the thickness of the particle. This is true for the cases of the solid-solution pathway. However, in the cases of phase separation, almost all Li$^+$ ions quickly equilibrate within the stable phase domains and only make concentration jumps at the phase boundary between the two phases (colors), which is confirmed by the *operando* observations that the net Li$^+$ ion flux through the edges concentrate at the phase boundaries without causing any color change in the stable domains. In another word, Li$^+$ ion reaction flux at the particle edges is equivalent to the internal Li$^+$ ion flux at the phase boundary that makes the phase boundary move. If the internal phase boundary does not move, no net electrochemical current will be seen on the edges of the particles. For consistent transport analysis, the flux that is normal to the particle edge but tangent to the phase boundary is less responsible for the movement of the phase boundary. Therefore, only the edges that are nearly parallel to the internal phase boundary should be counted as the effective reaction interface. We calculated the local working current densities or the interfacial current densities within each selected particle by dividing the local current with the associated active area, either the entire edge area for the solid-solution cases, or the phase boundary area for the phase-separation cases, as shown in **Figure 2 (a)-(d)**.

As shown by the schematic highlight in **Figure 2 (e)**, the single-particle level solid-solution behavior tends to induce concurrent reactions of all particles at the many-particle level. On the other hand, the single-particle level two-phase dynamics tends to activate just a few particles at a time[40,41], resulting in sequential reactions, which in turn yield a very high local working current density that can enable a solid-solution dynamics at the single-particle level. It is then not unexpected to discover that the true local current densities during phase separation are much higher than those in the solid-solution step (Figures 2 (a-d)). More specifically, nucleation events of a new phase that led to the emergence of phase boundaries always caused sudden current density spikes. The positions of these spikes during lithiation at 0.1 C current in **Figure 2**

**(a)** correspond to the particle-by-particle activation process. During the red-to-gold lithiation at 0.1 C current, more active phase-separating particles lowered the absolute current shared by each particle and therefore lowered the true local current densities (*0.1 – 0.2 mA cm$^{-2}$*) in individual particles shown in **Figure 2 (a)**. The overlapping current densities are consistent with the relatively concurrent reaction among particles. During delithiation at 0.1C current, the obtained true local current densities are similar to those during the lithiation, only that the SOC ranges changed due to hysteresis between lithiation and delithiation.[42,43] In the case of 1 C current in **Figures 2 (c)**, the solid-solution-like blue-to-red lithiation makes the quantification of phase boundary improbable. Only the red-to-gold lithiation exhibit phase-boundary-based high local current densities. True local current densities for the delithiation process consistently reflect the existence of phase boundaries and the sequential reaction process.

It is evident from the above observations that while the solid-solution mechanism led to a homogenous distribution of the external driving conditions on the entire domain, the phase-separation caused higher working current densities on limited active particles. These interfacial current densities were 2-3 orders of magnitude higher than the average current density calculated using the BET surface area (*9.424 m$^2$ g$^{-1}$*). Surprisingly, a ten-fold increase in the external current from 0.1 C to 1 C increased the working current densities only nearly 5 times. The high "working" current densities in the studied cases suggest a far-from-equilibrium (de)intercalation dynamics at both slow and fast galvanostatic conditions. These understandings also indicate that the process is not diffusion-limited,[28] and requires innovative theoretical analyses to uncover the underlying mechanism.

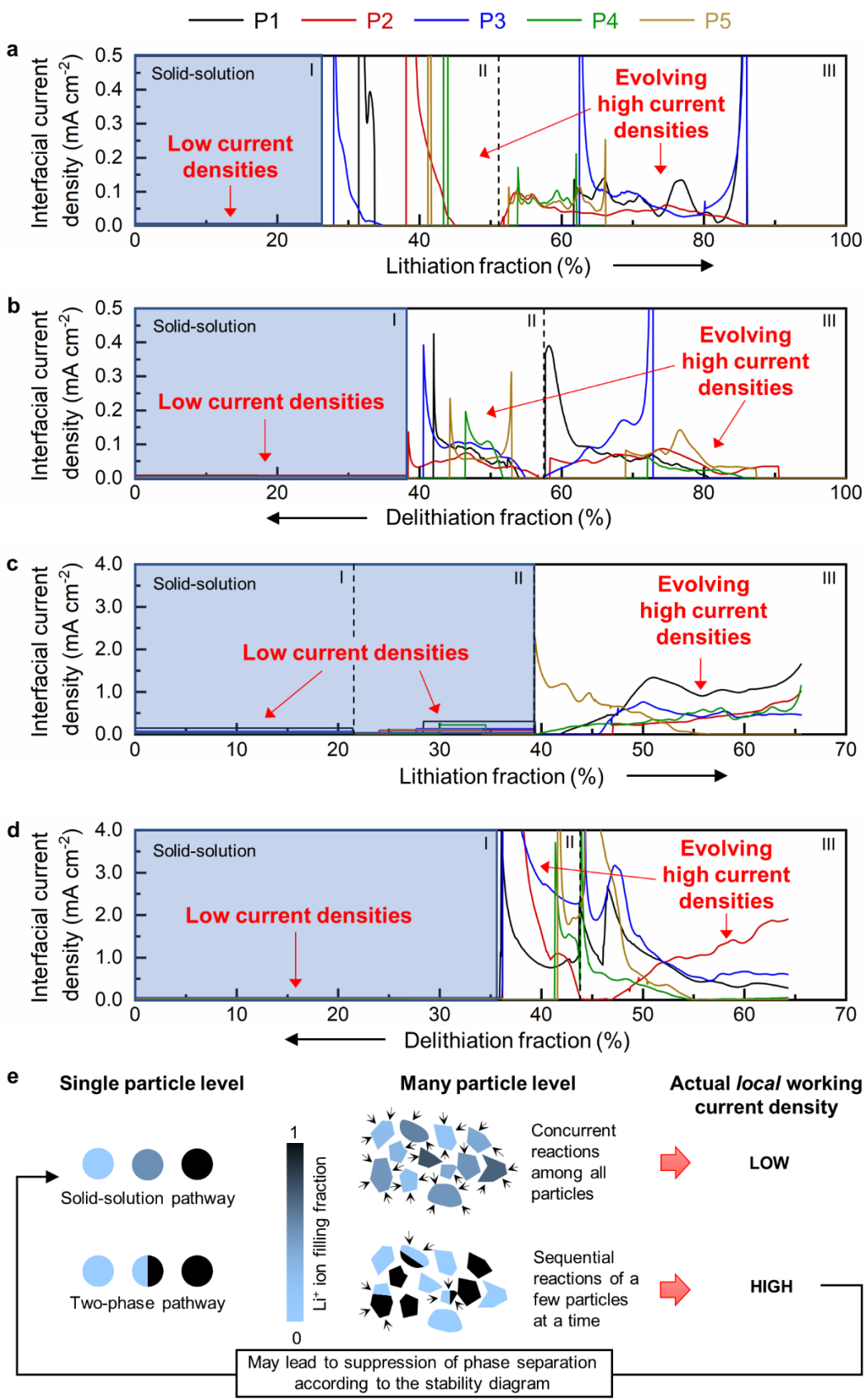

**Figure 2:** *Operando* **interfacial current densities during (de)lithiation in graphite under constant current.** *Operando* interfacial current densities in five selected particles estimated using the length of phase boundaries during (a) lithiation at 0.1 C current, (b) delithiation at 0.1 C current, (c) lithiation at 1 C current, and (d) delithiation at 1 C current. The solid-solution regions in these four panels (shaded in light blue in Step I and with the additional Step II for panel (c)) yielded low *local* interfacial current densities, while the two-phase regions yielded evolving but high local current densities. The distinct positions of the sharp peaks in Step II during lithiation at 0.1 C current indicate a particle-by-particle mechanism. The almost overlapping current density regions in the remaining phase-separation regimes show that the phase boundaries emerge simultaneously within these particles. (e) Schematic comparison of solid-solution pathway vs two-phase pathway at the single-particle level and many-particle levels, yielding different actual local current densities due to population of active particles[40] that in turn influence the single-particle dynamics.

## 2.3. Non-equilibrium thermodynamics of graphite

The thermodynamic origin of the electrochemical phase transformation has been well explained by the balance between the non-monotonic diffusional chemical potential and the applied electrochemical overpotential,[37,44] based on which it was predicted and verified in LiFePO$_4$ single particles[43] that phase separation, therefore the coexistence of two stable phases, can be suppressed by an external electrochemical driving force. Graphite, as another well-known phase separating material, has not been examined rigorously over the possibility of suppression of phase separation, despite a few attempts to explain the evolution of phase separation.[34,45,46]

Here, we performed a linear stability analysis[37] of the electrochemical phase transformation in a single graphite particle, by evaluating the dynamics of a linear perturbation, superimposed over a reaction-limited homogeneous system, but under various constant-current driven working conditions. An experimentally verified bilayer regular solution model,[33,47] in which the two-step diffusional chemical potential corresponding to the two phase transformation steps, was adopted here to investigate the stabilities. The bilayer regular solution model allows the filling fraction $\tilde{c}_1$ to vary between 0 and 1 while $\tilde{c}_2 \sim 0$ for blue-red phase transformation; once $\tilde{c}_1 \sim 1$, the second filling fraction $\tilde{c}_2$ starts to increase during red-gold phase transformation. As presented in **Supplementary Information Section 7**, our theoretical analysis provides the normalized growth speed of the perturbation with wave number k for the blue-red transformation,

$$\tilde{s}_1(\tilde{k}, \tilde{c}_1, \tilde{J}) = \left[2\tilde{\Omega}_a + 2\tilde{\Omega}_c \tilde{c}_2(1-\tilde{c}_2) - \frac{1}{\tilde{c}_1(1-\tilde{c}_1)} - \tilde{K}\tilde{k}^2\right]\sqrt{\frac{\tilde{J}^2}{4} + [F(\tilde{c}_1)]^2} + \left(\frac{F'(\tilde{c}_1)}{F(\tilde{c}_1)} + \frac{1}{2}\tilde{K}\tilde{k}^2\right)\tilde{J} \quad (1)$$

where $\tilde{\Omega}_a$ and $\tilde{\Omega}_c$ represents the regular solution coefficients for the intra-layer and inter-layer particle-vacancy interactions respectively; $\tilde{K} = V_S K/k_B T L^2$ is the dimensionless gradient penalty parameter dependent on the volume per intercalation site $V_S$, diffusion length $L$, Boltzmann constant $k_B$, and the room temperature $T$ taken as 298 K. Following a similar strategy, the stability equation of the growth speed during red-gold phase transformation is slightly different,

$$\tilde{s}_2(\tilde{k}, \tilde{c}_2, \tilde{J}) = \left[2\tilde{\Omega}_a + 2\tilde{\Omega}_c \tilde{c}_1(1-\tilde{c}_1) - \frac{1}{\tilde{c}_2(1-\tilde{c}_2)} - \tilde{K}\tilde{k}^2\right]\sqrt{\frac{\tilde{J}^2}{4} + 1} + \left(\frac{1}{2}\tilde{K}\tilde{k}^2\right)\tilde{J} \quad (2)$$

Apparently, the growth speeds $s_1$ and $s_2$ depend on $\tilde{J}$, which is the local interfacial current density $J$ scaled to the concentration-dependent exchange current density in the homogeneous state $J_0(\tilde{c}_i) = k_0 F(\tilde{c}_i)$ with $k_0$ being the rate constant and $F(\tilde{c}_i)$ being the normalized exchange current density function for respective phase transformations. The complete derivation of Equations (1) and (2) can be found in **Supplementary Information, Section S7**. It is important to note that the local exchange current density is an intrinsic interfacial property and needs to be determined experimentally in a self-consistent manner using the area of the active sites, instead of the total or apparent area. Based on our *operando* observations, image analyses, and impedance diagnosis,[28] the intrinsic exchange current densities for particles in the solid-solution and red-gold phase transformation states are relatively independent of the Li$^+$ ion filling fraction as *0.5 mA cm$^{-2}$* and *3 mA cm$^{-2}$*. The exchange current density for particles in the blue-red phase transformation, however, is sensitive to the filling fraction and appears as a skewed non-monotonic curve in **Figure S8**.

With the exchange current densities, we plotted in **Figure 3** the neutral and driven linear stability boundaries, $\tilde{s}_{max}(\tilde{k},\tilde{c}_1,\tilde{J}) = 0$ and $\tilde{s}_{max}(2\pi,\tilde{c}_1,\tilde{J}) = |\tilde{J}|$, using the most unstable mode $\tilde{k} = 2\pi$. The stability boundary for the blue-red phase transformation in Figure 3 (a) has a maximum, resembling the diagram for LiFePO$_4$.[37] The relatively constant exchange current density for the red-gold phase transformation, however, leads to different stability boundaries that do not close to reaching a maximum, as shown in Figure 3 (b).[43]

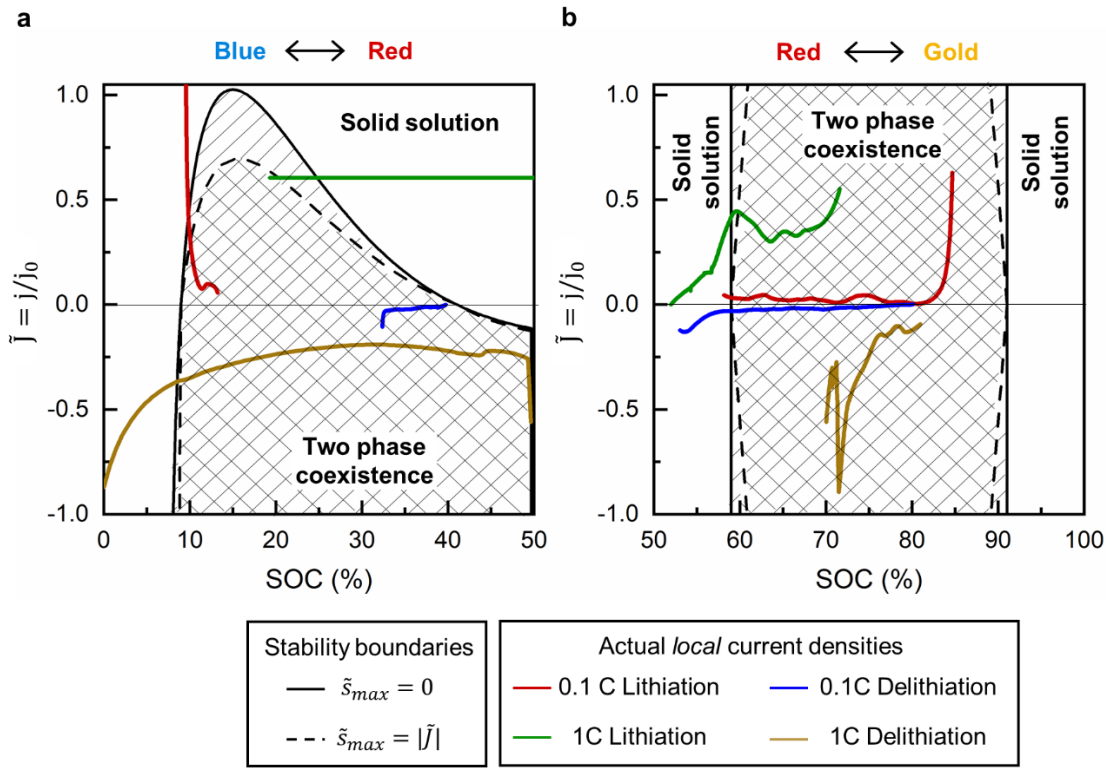

**Figure 3: Comparison of *operando* current densities and the linear stability diagram for graphite single particles.** Linear stability diagram for (a) blue to red, and (b) red to gold phase transformations during both lithiation and delithiation at both the galvanostatic conditions. The experimental *operando* current densities (Steps II and III) lie within the domains of two-phase coexistence in all the cases except the blue-red lithiation process at 1 C current (green solid curve in the SOC range of 20% to 50% in panel a). This observation is consistent with the visual examination of the (de)lithiation process shown in Figure 1. Current density data displayed here are from particle P1. The stability diagram plays a pivotal role in the dynamic interplay between the nonequilibrium thermodynamics and electrochemical kinetics, which can be understood by a flow chart in Figure S10.

The linear stability diagram acts as a "nonequilibrium electrochemical phase diagram" generated from the interplay between the nonequilibrium thermodynamics of the materials and the exchange current densities at the materials interface. The stability diagram is a rigorous quantitative summary of the dynamic interplay. As shown in the flowchart (Figure S10), conventional understandings following the classic thermodynamics would consider that a near-equilibrium excitation will only induce near-equilibrium dynamics. Therefore, a low current applied to phase-transforming electrodes would induce two-phase coexistence in every single particle. However, as we proved here in complex phase-transforming electrodes like graphite, phase separation will lead to highly localized current density, that may lead to the suppression of phase separation and thereafter the solid-solution dynamics, but only if a SOC-dependent non-monotonic exchange current density is in place. As discussed above, the phase separation can then be suppressed only in the blue-red phase transformation regime, owing to a skewed concentration-dependent exchange current density. Due to the non-monotonic stability curves for the blue-red phase transformation during the lithiation process, there exist critical current density $\tilde{J}_{BR}^c \sim 1.02$ beyond which the phase transformation would occur via the solid-solution pathway. On the other hand, the diverging unstable region for the red-gold phase transformation during both the lithiation process and the delithiation process suggests that the respective phase separations will always be triggered, irrespective of the current densities. Take particle P1 as an example, we overlay the *operando* interfacial current densities from **Figure 2** on the stability diagram **Figure 3**, and observed that only the blue-red transformation during lithiation at 1 C current lied close to the critical current density $\tilde{J}_{BR}^c$, where a dynamic solid-solution pathway becomes possible. The *operando* current densities during the other phase transformation processes, however, lied within the phase-separation domains for all the phase transformations. Similar results were obtained for the other four particles and are shown in **Figure S9** in the Supplementary Information. It should be noted that the regular solution model spans across the entire SOC range from 0%, to 100% SOCs, while the actual phase-transformation in our electrodes occurs within different SOC ranges shown in **Table S1** in **Supplementary Information**. To better compare them, the *operando* interfacial current densities have been scaled (only horizontally along the SOC axis) to match the phase-transformation SOC ranges of the regular solution model. This theoretical understanding offers a consistent explanation of the visually obtained local current densities during phase transformations and indicates that the dynamics is indeed controlled by the *operando* electrochemical flux, and not the solid-state diffusion.

## 2.4. Simulations of mesoscale many-particle dynamics

The single-particle analyses necessitate the incorporation of non-equilibrium materials thermodynamics into electrochemical models[37,44,48] toward the holistic design of battery materials and electrodes. Here, we adopt the multiphase porous electrode theory (MPET) developed by the Bazant group,[49] which not only incorporated materials thermodynamics for phase transformation, but also can predict the transport and polarization processes through the electrode that can be modeled by the classic porous electrode theory pioneered by J. Newman.[50,51] With the consistent kinetic parameters, i.e., Li$^+$ ion diffusion coefficient $D_{Li}$ and exchange current density $j_0$, now extracted from our *operando* experiments, we modeled 200 thin flake-like rectangular graphite particles to simulate the many-particle dynamics in practical surroundings. The cluster of particles has a fixed width (*8 μm*) and thickness (*0.5 μm*) with the length dimension lognormally distributed with a mean and standard deviation of *8.13 μm* and *2.61 μm*, respectively (same as experiments). The model framework adopts a two-parameter Cahn-Hilliard reaction theory where the equilibrium voltage curve is represented by a periodic two-layer regular solution model[33] and the intercalation follows Marcus kinetics.[52] The reduced 1D simulation with surface reaction boundary condition applied on both ends of the particle was performed. To replicate the true local dynamics, we employ the values of the Li$^+$ ion diffusion coefficient in graphite $D_{Li}$ and the rate constant $k_0$ evaluated from the *operando* interfacial current densities in our recent work.[28] The exact values of $D_{Li}$ are *3.25 ×10$^{-7}$ cm$^2$ s$^{-1}$* and *1.25 ×10$^{-7}$ cm$^2$ s$^{-1}$* and $k_0$ are *4 mA cm$^{-2}$* and *3 mA cm$^{-2}$*, respectively for blue-red and red-gold phase transformations with the Cahn-Hilliard gradient penalty parameter *K* required in the equilibrium voltage calculation as *9×10$^{-7}$ J/m* for the phase transformations.[38] The details of the MPET model can be found in multiple reports by Bazant and collaborators.[33,36,49,53] The key governing equations used in the model are presented in the **Supplementary Information Section 9**.

As suggested in Section 2.2, we rigorously studied the phase transformation by evaluating the progression of the area fractions of the stable phases in the lognormally distributed particles. With the length dimension sorted in ascending order and assigned respective numerical IDs as shown in **Figure 4**, we identified the phases based on the concentration gradient at the phase boundaries, similar to **Figure S11** for Particle ID #100. Here, the solid-solution processes, which include both the empty (grey) to Stage 3 (blue) and Stage 3 (blue) to Stage 2 (red) lithiation at 1C (but not at 0.1C), would appear blue. In cases of phase separation, the vertical line/region corresponding to each particle ID will show the coexistence of the two colors in accordance with the respective phases. The lithiation at 0.1 C current (**Figures 4 (a-d)**) began with homogeneous intercalation within all particles until the battery voltage reaches the spinodal point (~4% SOC), which triggered the blue-red transformation.[37] The smaller particles showed the blue-red phase boundaries first and became completely red before inducing the phase transformation in new particles. This particle-by-particle filling mechanism continued in the electrode without any appearance of the gold color, except in a small overlapping SOC range of 50% - 60%. Multiple particles activated during the red-gold phase transformation and possessed phase boundaries simultaneously. But the smaller particles tend to completely transform into gold earlier than the larger ones. The lithiation at 1 C current (**Figures 4 (e-h)**), however, mapped blue color in all the particles up to ~34% SOC, indicating that the blue-red phase transformation continued the solid-solution behavior even after all the particles reached stage 3. While some smaller particles exhibited the blue-red phase boundaries, the gold color appeared at ~37% SOC. The red-gold phase transformation then continued similar to the lithiation at 0.1 C current with significant remaining phase boundaries at the end, due to a reduced electrode capacity at the cut-off voltage (~65%). The delithiation process at both 0.1 C and 1 C currents, however, exhibited gold-red and red-blue phase transformations via the respective phase boundaries, as shown in **Figure S12**. The model response was consistent with the experimental *operando* observations, confirming its validity. The accurate prediction of the phase transformation pathways at both low and high currents provides a validation of the kinetic parameters and also demonstrates the predictive capability of MPET to understand the dynamics of the battery materials.

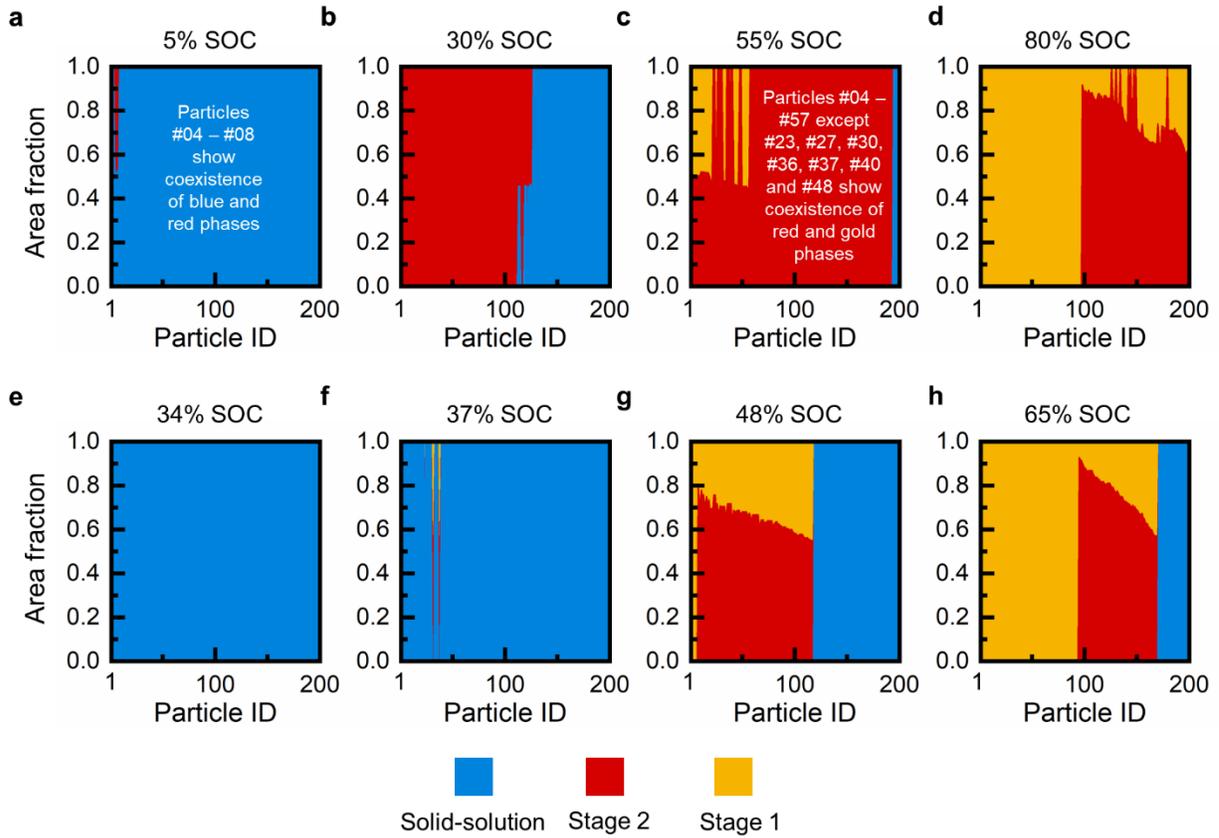

**Figure 4: Theoretical predictions of area fraction of the stable phases using MPET with 200 lognormally distributed particles.** Area fraction of stages 3, 2 and 1 at four global SOCs at **(a-d)** 0.1 C current, and **(e-h)** 1C current. The increasing particle IDs signify increasing particle lengths. At 0.1 C current, the blue-red phase boundaries appear initially in smaller particles. The new phase boundaries begin only when the phase transformation in the earlier particles is finished. On the other hand, the solid solution behavior continues up to ~34% at 1 C current which means that the blue-red phase transformation also occurs almost via the solid-solution mechanism. Once almost all particles are in stage 2, the red-gold transformation occurs along the phase boundaries at both currents.

In addition to the qualitative consistency, the simulations allow evaluation of the instantaneous total interface length within the simplified ideal rectangular particles, using the concentration gradient at the phase boundaries. Based on the 1D simulation, the phase boundaries emerged parallel to the length dimension of the particles. Thus, the total length of the phase boundaries was equal to the collective length of the active particles and could be extrapolated to the total volume of graphite electrode used in the experiments. These simulated spatiotemporal responses of phase boundaries in the considered galvanostatic conditions, during both lithiation and delithiation, despite notable deviations, were in the same orders of magnitude with the *operando* observations, as shown in **Figure 5 (a-b)** and **Figure S13**, respectively. The widened curves and shifted peaks in the simulated lengths of the phase boundaries, as compared to the image analyses, are mainly due to the inconsistency introduced by the simple 1D particles, in which the ideal 1D phase boundaries cannot reflect the 2D results perfectly. While the numerical challenges in simulating hundreds of realistic 2D particles will be addressed in our future works, the comparison shown in Figure 5 is still deemed meaningful, as it closes the loop of our analysis: using kinetic parameters ($D_{Li}$ and $k_0$) extracted from true local electrochemical processes and incorporating nonequilibrium materials thermodynamics enable mathematical simulations that can capture the spatiotemporal heterogeneities at multiple scales in complex phase-transforming electrodes.

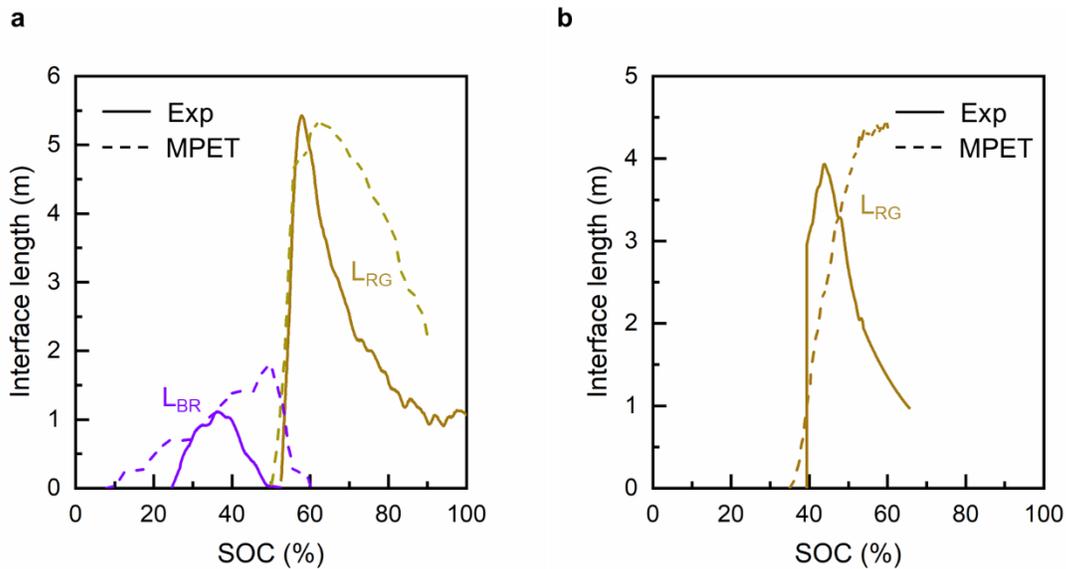

**Figure 5: Comparison between lengths of phase boundaries extracted from direct image analysis (solid lines) and that simulated using MPET (dashed lines) at (a) 0.1C current, and (b) 1C current.** The blue-red transformation begins earlier in the simulation than in experiments because the simulation triggers phase-transformation when the Li$^+$ ion concentration reaches a spinodal point (~4% SOC) in the regular solution equilibrium voltage curve.[37]

## 3. Discussion

Quantifications of the phase boundaries and hence the *operando* interfacial current densities provide a consistent understanding of the active reaction fronts and the local electrochemical activity in the phase-transforming porous electrodes. A three-orders-of-magnitude higher interfacial current density than the averaged one, at a low current of 0.1 C, suggests a far-from-equilibrium dynamics as opposed to widely known near-equilibrium process, and necessitates special attention in interpreting experimental data. Moreover, a comparable magnitude of interfacial current densities at 0.1 C and 1 C currents highlights that the local reaction flux remains almost unchanged on increasing the total current, as the interfacial reaction appears to be coordinated and modulated by the phase transformation dynamics among many particles. Despite the similar orders of magnitude of the interfacial current densities at low and high currents, the intercalation mechanism may be different, owing to the intrinsic exchange current density affected by the local filling fraction.[28] Our results show that only the blue-red phase transformation during the lithiation process can be suppressed at a high current, while red-gold phase transformation would always occur via phase-separation, shown in the electrochemical stability diagram in **Figure 3**. However, our *operando* experiments at 1.5 C and 2 C currents show that in real battery operation, the much larger overpotential resulted in the incomplete charge/discharge at the cut-off voltages, that the capacity obtained at such high currents would be insufficient to start the red-gold transformation, as shown in **Figure S1**. Therefore, the spatiotemporal heterogeneities seem to dampen at high currents in our thin graphite electrodes shown in **Figure S14**.

There are two experimental aspects pertinent to this study, enabling us in successful analysis of the spatiotemporal heterogeneities. Firstly, we ensured a low thickness of the electrode (~ 5 μm) to avoid the occurrence of the reaction heterogeneities along the thickness due to the concentration gradient in the electrolyte.[6,15,23] A consistent comparison between the total capacities obtained from the electrochemical experiments and the direct image analyses, confirms a similar level of reaction heterogeneities in the entire electrode. Secondly, we used flake-type graphite particles to allow easy in-focus imaging of hundreds of particles with high resolution at single-particle level. The underlying conclusions, based on the fundamental phase transformation characteristics, apply to other types of graphite such as mesocarbon microbeads and synthetic graphites,[6,54] in which clear phase transformation and reaction heterogeneities were observed.

A recent scaling analysis suggests that the Li$^+$ ion intercalation in large graphite particles of 50 μm[38] and 400 μm[55] is diffusion-controlled, but according to our latest observations,[28] the intercalation process in our graphite flakes at mesoscale is governed by phase-transformation. For this reason, the generalized electroanalytical methods, independent of the intercalation and rate-controlling mechanisms, should be used to estimate the consistent values of $D_{Li}$ and $j_0$.[28,56,57] The other way can be to utilize the detailed phase-field models with non-equilibrium thermodynamics, such as MPET, to estimate these kinetic parameters, which are the only fitting variables in them. Once the phase boundaries and the true working current densities in the phase-transforming materials are identified, and the phase-field models are calibrated with the *operando* observations, these models can simulate the local electrochemical behavior under practical conditions. The one obvious modification in the MPET would be to extend the simulation to two dimensions, which would accurately replicate the intercalation process in the layered phase-transforming materials such as graphite. With this modification, the physics-based robust models such as MPET can be a significant addition to the battery management systems in the commercial batteries for an accurate estimation of the actual SOCs, which is still a persistent issue for the battery manufacturers.

## 4. Conclusion

We have utilized a simple benchtop *operando* setup to observe the spatiotemporal heterogeneities under various constant current densities. Unlike other well-known phase-transforming battery materials, not all phase separation processes in graphite can be suppressed by a high current. By comparing the true local current density of a single particle from direct image analysis with the linear stability diagram of electrochemical phase transformation, it is now clear that whether an ion-intercalation-induced phase transformation can be suppressed or not, depends on the intrinsic interfacial property. A concentration-dependent non-monotonic exchange current density is the premise for the suppression of phase separation, via the autoinhibitory pathway.[58] Beyond the consistent single particle analysis, a many-particle model was developed to understand the mesoscale behavior of the electrode, by using the MPET code. The results consistently replicate the mesoscale behavior of the $Li^+$ ion intercalation and deintercalation into graphite porous electrodes and support a coherent thermodynamics-incorporated model that can simulate not only the intra-particle but also the inter-particle dynamics in the practical battery systems. The findings presented in this study clearly reveal a strong connection between the spatiotemporal heterogeneities and the mechanism of phase transformations in the electrodes, and its incorporation in the electrodes design can assist in their high-rate performance.

## 5. Experimental Section

**Thin electrode preparation:** We formed a homogeneous graphite slurry by dissolving graphite flakes (7-10 μm, 99%, Alfa-Aesar), PVdF binder (>99.5%, MTI Corp) and conductive acetylene black powder (35-40 nm, MTI Corp) in the ratio 88:10:2, in 1-methyl-2-pyrrolidone (NMP, 99.5%, Sigma Aldrich). This slurry was coated onto separator film by the doctor-blade method. We, then, dried the electrodes at 60°C to remove the NMP, punched out Φ8 mm electrodes and kept them under vacuum at 70°C for 12 hours to remove the residual moisture. The active material loading, and electrode thickness were 0.7 mg cm$^{-2}$ and 5 μm, respectively, shown in Figure S14.

*Operando* **setup and experiments.** We assembled a graphite half-cell with a Li anode, a glass-fiber separator, and 1M LiPF$_6$ in EC:DMC (50:50 v/v) in a standard 2032 coin cell with a Φ2 mm hole on the top, in an Ar-filled glovebox. We sealed the viewing hole with a 5 x 5 mm glass window using epoxy to seal the cell to view the graphite flakes under the optical microscope Olympus BX53M microscope with objective 50x for *operando* observation. We ran five formation cycles at 0.25 C between cut-off voltages 1.5V and 0.4 mV. We then, performed a galvanostatic charge and discharge at 0.1 C and 1 C currents, and captured the frames every 20s and 2s time intervals, respectively. All the acquired digital photos were processed using ImageJ to quantify the colored regions. The detailed description of the procedures is mentioned below.

**Color thresholds for area quantification.** We used the built-in color threshold method of ImageJ identify the blue, red and gold colors based on the Hue, Saturation and Brightness in the frames captured in the *operando* galvanostatic experiments. The brightness difference of blue, red and gold colors caused some smaller particles (< 2 μm), which were not clearly visible when they were blue or red, to get illuminated upon turning into gold color. Since ImageJ auto-selects all the non-black regions/graphite flakes according to the brightness, this illumination resulted to select more area when the particles entered Step III (or Stage 1). The difference in the selected area across all the images was always under 10%, resulting in slight inconsistency while calculating the area fraction of different phases. To avoid this inconsistency, the same sampling region was defined based on the images with all red-colored particles while converting the surrounding areas black. The black mask constituted all the non-particles regions and the unreacted regions obtained from 100% SOC image. This final black mask was applied to all the images. Within the sampling regions, a fixed range of hue was used to select similar colored regions (Red: 0 – 24, Gold: 24 – 44, and Blue: 44 – 255) while maintaining the same range of saturation and brightness. The above criteria were applied to all the digital images with the help of an ImageJ script. A variation of ±10% in the selected thresholds results in a deviation of only ±0% – 5% in the selected areas of blue, red and gold.

**Charge conservation calibration.** We first converted the blue, red, and gold colors into the standard RGB colors using our segmentation algorithm in ImageJ, explained above, and determined the evolution of area fractions ($a_i$) of these colors in the overall view (**Figure S3**). Since these colors represent stable phases of the lithiated graphite, they have a fixed SOC with respect to the total electrode capacity, associated with them. During phase transformation at near-equilibrium conditions, the transformed particles wait until the entire electrode reaches the same state, making the SOC of that particular state same as the global SOC. Hence, we determined the SOCs of the blue ($x_B$), red ($x_R$) and gold ($x_G$) colors as 26%, 55% and 100% under low galvanostatic conditions (C/72 current). We used these SOC values for the validation of our color segmentation algorithm with the electrochemical response, via charge conservation within the electrode, $\sum_i q_i A_i(t) = q_T(t) A_T$, where $i$ represents Blue, Red, and Gold, $q_i$ is the areal capacity of the $i^{th}$ color and can be calculated from the SOC of the $i^{th}$ color (estimated above) and the theoretical areal capacity of the material ($q_o$), $q_i = x_i q_o$, $A_i$ is the area covered by $i^{th}$ color, $q_T$ is the areal capacity of the electrode and $A_T$ is the total surface area of particles in the electrode. The above equation can, then be transformed into $\sum_i x_i a_i(t) = x_T(t)$ where $a_i$ is the area fraction of the $i^{th}$ color, and $x_T$ is the global SOC of the electrode. For known area fractions of stable phases, a phase-transforming material should inherently follow this equation. Since we only observed a $100 \times 100 \mu m$ window under the optical microscope, the validity of the above equation for the observed region confirms that the analysis can be confidently extrapolated to the entire electrode (**Figure S4**).

**Determination of the effective interfacial area:** For a shape with two colors, we used ImageJ to find the perimeter covered by each color, and the outer perimeter of the shape, which together can be solved for the length of the interface. In our case, particles existed in three different states at a time. To calculate the length of the interface, for instance, the Blue – Red interface, we relied on the fact that the phase transformation in graphite can only occur in one order: Stage 3 to Stage 2 to Stage 1. We converted all the green regions in the transformed RGB images, to the standard red color, thus eliminating all Red – Gold interfaces. This enabled us to find the length of the Blue – Red interface using the above methodology from the following equation, $L_{BR} = (l_{Blue} + l_{Red} - l_{particles})/2$. Similarly, we calculated the length of the Red – Gold interface by converting all the blue regions to the standard red and applying following equation, $L_{RG} = (l_{Red} + l_{Gold} - l_{particles})/2$ where $L_{BR}$ and $L_{RG}$ are the lengths of Blue – Red and Red – Gold interfaces respectively, $l_{Blue}$, $l_{Red}$ and $l_{Gold}$ are the perimeters of the blue, red and gold regions in the corresponding transformed images, and $l_{particles}$ is the outer perimeter of all the particles within the viewing frame.

Considering disc-shaped flakes with an average diameter 8µm and thickness 0.5 µm, our ~5 µm thick electrode constituted ~$10^7$ particles with 10 layers stacked over each other. On making a statistical assumption that all layers were similar, we calculated the active area by multiplying the total length of phase boundaries with the electrode thickness.

**Acknowledgment**

P.B. acknowledges the support from a National Science Foundation grant (Award No. 2044932), and the faculty startup support from Washington University in St. Louis. The materials characterization experiments were partially supported by IMSE (Institute of Materials Science and Engineering) and by a grant from InCEES (International Center for Energy, Environment and Sustainability) at Washington University in Saint Louis.

**Author contributions**

P.B. conceived and supervised the study. S.A. and P.B. designed the experiments. S.A. performed the experiments, carried out the analysis. S.A and P.B. wrote and revised the manuscript.